**Abstract**
This study aims to reveal different varieties of capitalism and to uncover new patterns of development that emerged between 2010 and 2020. A hybrid model is applied that quantifies three pillars of development (Future - F, Outside - O, Inside - I) using supply-side and demand-side indicators that measure norms, institutions, and policies. Investigating 34 OECD members, this study describes five varieties of capitalism: traditional, dualistic, government-led, open market-based, and human capital-based models. It is suggested that the most significant cut-off point in the development of OECD economies in this period was along the green growth dimension, where European countries with a tradition in coordinated markets outperform the rest. Using Israel and Estonia as an example, it is also suggested that institutional and policy changes that enhance the quality of governance and make coordination more effective are the way out of the middle-income trap.

Keywords: development path; economic policy; green growth; institutions; middle-income trap; OECD; varieties of capitalism

**Introduction**

The concept of path dependence first appeared in economics in the 1980s (Gigante, 2016) and has spread very quickly in micro and macro applications/theories. Lying behind this massive increase in popularity was the growing significance of historical, geographical, and cultural contexts in economic analysis. In development economics, the main message of path dependence is that the development path of different economies cannot be described using a universal mathematical model. This, of course, has always been the starting point in the theory of comparative economic systems. Comparative economics first focused on the differences between centrally planned and market economies, but following the Second World War it was pointed out that even Western market economies do not follow the same path (Shonfield, 1965). This area of comparative political economy has been rejuvenated in the 21st century as the analysis of varieties of capitalism (Hall and Soskice, 2001).

The varieties of capitalism were generated by path dependence, and their characteristics are relatively stable as they are strongly influenced by traditions. It is unlikely that a country will switch from one market model to another one in a matter of one or two decades. Some change is still expected, though, as new challenges (the most discussed such macro trends are globalisation, digitisation, and the climate (Anděl et al., 2022)) force policy changes, and these can ultimately lead to diverging or converging paths in the varieties of capitalism. It is commonly (but not undisputedly, see Haagh, 2019) accepted by the researchers of this field that there is no such thing as a best version of all the varieties of capitalism (in other words: many different versions can generate high levels of wellbeing), but successful policy responses given to new challenges become popular, and they have an obvious impact on the economic policy choices of most countries.

The question asked in this study is whether this chain of (1) new challenges, (2) policy responses and (3) adaptation processes can lead to new patterns in the varieties of capitalism, and if yes, what these patterns are. To answer these questions, this study adopts the FOI model (Bartha and Gubik, 2014) originally developed at the University of Miskolc in the early 2010s. This paper contributes to the research on varieties of capitalism in the following three areas:

1. By adopting the FOI model, and distinguishing between the future, outside, and inside dimensions of development, it offers a novel interpretation of the varieties of capitalism.

2. By comparing the 2010 and 2020 FOI values calculated for the OECD countries, possible shifts in the development paths of the most developed economies can be detected.
3. By looking at the possible influencers of the development path shifts, it can offer explanations on what new patterns are emerging among the OECD economies.

The rest of this study is structured as follows: the first chapter presents a review of the literature on the varieties of capitalism; the second part presents the methods of the study, focusing on the FOI model; the third part presents the results; finally, the paper offers a discussion of results.

**Literature review on varieties of capitalism**

The comparative political economy of the 1980s separated mixed market economies into three distinct groups according to the role the state plays in coordinating the market economy. Katzenstein (1985) discusses the comparative advantage of small open economies in Europe, and by focusing on industrial and income policy specifics argues that this group of countries (such as Austria, Denmark, Netherlands, and Sweden) forms the third model of mixed economies: corporatism (the other two being statism, e.g., France or Japan, and liberalism, e.g., the USA). Zysman (1983) analyses national financial systems, but his conclusion is very similar to Katzestein's. He also outlines three varieties of financial capitalism: government-led credit-based, bank credit-based and capital market-based. There are strong parallels between the two classifications. The government-led credit-based system corresponds to the statist model, the bank-credit-based model corresponds to corporatism, and the capital market-based model is similar to liberalism. These three-way classifications faded away in the 1990s. Economic reforms of the decade, for example, made industrial policy and government-led credit allocation outdated.

The concept of national competitiveness brought new life to the literature of varieties of capitalism. Hall and Soskice, in their seminal work (2001), based their argument on supply-side competitiveness factors, and outlined two major models: liberal and coordinated market economies. This new approach highlights the so-called comparative institutional advantage, a phenomenon that is generated by the interaction of certain institutions in the national economic system, manifesting as a distinct set of patterns characterising the economic performance and specialisation of different sectors of the economy. Hall and Soskice (2001) originally focused on the institutions influencing the division of tasks among governments, corporations, and labour, and paid special attention to factors such as the internal structure of corporations, company and labour relations, and financial and education systems. The liberal and coordinated version of market economies represents different sets of comparative institutional advantages, which are reflected in the structure of the economy. It can manifest itself in innovation performance as well, and this proposition has generated a great deal of empirical papers (Witt and Jackson, 2016 not only offer such an analysis, but they also provide a substantial review of the literature). The results related to innovation and the two varieties of capitalism were controversial. David Soskice in his later work investigated national innovation systems in great detail and concluded that the USA should be removed from the group of liberal market economies, hence, he created a third category. He argues that the United States is in a class of its own because it plays such a unique role in generating disruptive innovations (Soskice, 2022). Just as Hall and Soskice, Amable (2003) has also based his approach on the characteristics of the input and output market, the financial system and, additionally, on the welfare and the education system. His approach yielded five different varieties of capitalism: market-based Anglo-Saxon, Asian, continental (European), social democrat and Mediterranean. An important part of Amable's reasoning is that it is not only path dependence that leads to the different models, but rather the fundamental factor is the complementarity among certain institutions. Once a policy choice is made in the welfare, education, or financial system, the newly established institution will attract and enhance other institutional solutions, and this is how specific mixes are formed.

Schmidt (2005) claims that the new approach to the varieties of capitalism introduced by Hall and Soskice obscures the important role of the state in coordinating the economy. She insists that the method of state coordination is crucial in this field. Even though there have been shifts in the theory and practice of market-based state coordination, three models, namely the liberal (Anglo-Saxon), enabling (German, Dutch) and enhancing (French, Italian) models, are still clearly distinguishable according to Schmidt. The enhancing model is the one that is added to the original liberal and coordinated dichotomy; it is characterised by strong government intervention. Similarly to Schmidt, Kim and Kim (2021) also introduce state intervention as a pillar of classifying the varieties of capitalism.

Farkas (2016) focuses on the members of the European Union using an approach similar to that of Hall and Soskice. She conducted a cluster analysis and identified four groups: North-western, Nordic, Mediterranean, and Central-Eastern European. An interesting aspect of these empirical results is that Germany, the UK, and France are all part of the same cluster, while previous efforts have placed these major European economies into two or three different varieties of capitalism.

The financial crisis of 2007-2008 has put the demand side approach back on the map of mainstream economics, and so it was introduced into comparative economics as well. Baccaro and Pontusson (2016) examined the growth patterns of four European countries and distinguished between export-led and consumption-led models. A key feature of the latter model is that consumption is fuelled by private sector borrowing; therefore, the consumption-led model can also be interpreted as a borrowing-led path. Therefore, a parallel can be drawn between the liberal and coordinated models (defined in the supply-side approach) and the export-led and consumption-led models (defined in the demand-side approach). Baccaro and Pontusson remark that the number of possible development paths can be much higher than that concluded using the supply-side approach. This is confirmed by the work of Hein et al. (2021), which distinguishes between a weak and a mercantilist version of export-led regimes, and a domestic demand-led and a private demand boom version of debt-led regimes.

A different demand-side approach is presented by Morlin et al. (2022). They separate four different categories of autonomous demand (government expenditure, export, private investment, and debt-financed consumption) and use a method based on the supermultiplier. Their study concludes that export and government expenditure are the two dominant growth factors in Europe. The USA on the other hand seems more equally balanced. According to Morlin et al.'s calculations, if there is a decline in one of the categories of autonomous demand, it is in the USA where it is most likely that another category can step in and play the role of the growth engine.

Kohler and Stockhammer (2022) conducted an empirical study involving 30 OECD countries in the pre-and post-2008 period. They focus on demand-side growth factors, namely, real estate prices (measuring the extent of debt-led consumption), the structural budget deficit (government expenditure), the real-effective exchange rate (wage level), and the economic complexity of exports.

Another study suggesting a step away from the microeconomic framework was conducted by Blyth and Matthijs (2017). They argue that macroeconomic interventions should be considered when modelling the varieties of capitalism, and they suggest moving the foundations of comparative analysis to so-called macroeconomic regimes. Macroeconomic regimes are defined by the main target of a country's macroeconomic policy, and the two most common of these regimes are the ones focusing on employment (Keynesian) and price stability (neoliberal). Stockhammer (2022) gives a very detailed description of how the focus of comparative political economy has kept shifting from supply side (micro foundation) approaches to demand side, macroeconomic policy approaches, and back.

Development economics has traditionally emphasized the importance of the international context in development. Concepts such as the equal or unequal nature of trade relations (Balogh, 1963), one-sided specialisation (Myrdal, 1957; Prebisch, 1964; Singer, 1964), or the core-periphery dichotomy (Wallerstein, 1974) have contributed to the discussion of the topic. One valid criticism of the varieties of capitalism literature is that it assumes that the investigated national economies are not influenced by their international environment. This criticism has led to the emergence of the variegated capitalism literature (Peck and Theodore, 2007; Streeck, 2010; Jessop, 2014). The variegated capitalism line of literature concentrates on the interdependencies among national economies. Different varieties coexist and co-evolve, and dominant economies exert influence on the development model of others (Jessop, 2014). Variegation therefore emphasizes the interconnected and hierarchical nature of the world economy, which creates an environment within which development is combined and uneven (Dale and Unkovski-Korica, 2023). Therefore, the variegated approach focuses on individual cases, and the arguments are elaborated through qualitative case studies to reflect the complexity of the world economy and the institutional-cultural environment (Peck and Theodore, 2007; Peck, 2023).

This study uses the FOI model to investigate the varieties of capitalism within the OECD. When it was first introduced, it defined four different varieties of capitalism (the original study called them development paths): traditional-outdated structure; external resource focus; internal safe haven focus; and knowledge generation focus (Bartha and S. Gubik, 2013). The main features of this analytical tool are presented in the next chapter.

**Materials and methods – the FOI model**

The FOI model defines three pillars of development: F (Future); O (Outside), and I (Inside) factors (Bartha and S. Gubik, 2013). The Future potential measures factors that are crucial for the future competitiveness of the economy; the Outside potential captures factors determining the current world market position of the economy; and the Inside potential describes factors influencing the country's current level of well-being.

Table 1. Components of the F, O, and I potential, and the factors used to measure them

| Pot. | Component | Factor |
|---|---|---|
| F | Social responsibility | Solability Global Sustainable Competitiveness Index |
| F | Work ethic | WEF Cooperation in labour-employer relations |
| F | Energy efficiency | WEF Electricity supply quality % of output |
| F | Education expenditure | OECD Total expenditure on educational institutions as a percentage of GDP |
| F | Ageing of society | OECD 65 and above population, % of population |
| F | Renewable energy | OECD Renewable energy, % of primary energy supply |
| F | Environmental sustainability | GFN Ecological footprint |
| F | R&D potential | WEF R&D expenditures % GDP<br>WEF Patent applications per million population |
| F | Efficiency of the education system | OECD-PISA Not low achievers in reading, math & science, 15-year-olds |
| O | Trade openness | OECD (Exports+Imports)/GDP*2 |
| O | Country risk | TE Country credit rating |

| O | Stability of the financial sector | WEF Soundness of banks |
|---|---|---|
| O | Exchange rate stability | IMF 2017/2019 SDR variance |
| O | Language skills | ETS.ORG TOEFL scores |
| I | Government efficiency | WEF Budget transparency |
| I | Social wellbeing | OECD Better life index |
| I | Tax burden | IMF General government revenue, per cent of GDP |
| I | Pension system | OECD Assets in pension funds and all retirement vehicles, % of GDP |
| I | Level of development | IMF PPP GDP per capita |
| I | Growth | IMF Real GDP change |
| I | Availability of capital | WEF Financing of SMEs |
| I | Labour market flexibility | WEF Labour market Flexibility |
| I | Employment | OECD Labour force, % of population |
| I | Skilled labour | WEF Ease of finding skilled employees |

Source: own elaboration based on (Bartha and S. Gubik, 2013)

The FOI model represents a hybrid approach to the varieties of capitalism. It includes components that characterise norms (e.g., work ethics or social responsibility), and so they can only be changed very slowly (it takes decades to observe a change in them). Many other components describe institutions (e.g., education system, government regulation), and some of them are related to government policies (e.g., sectorial budget expenditures). These latter ones can change rather quickly. Most of the components included in the FOI model are supply-side factors, but there are also demand-side elements (e.g., exports, and government expenditure). From the focus areas introduced by Hall and Soskice (2001), the FOI model discusses the characteristics of company operations, labour relations, and the financial and education system. Welfare issues touched upon by Amable (2003) are also included in the FOI approach (through such components as social well-being or the pension system). Some components reflect on the international interdependencies (trade openness, country risk, exchange rate stability) highlighted by the variegated capitalism literature.

The FOI approach has four potential advantages that can justify its application. The components used to calculate the three indices were picked based on a thorough review of the development economics literature. Bartha and Gubik (2014) sort the development factors into internal and external groups and pick indicators through which they can be included in the FOI analysis. These development factors include the availability of capital (Harrod, 1948), factor productivity (Solow, 1956), entrepreneurial and innovation activity (Aghion and Bircan, 2017; McClelland, 1953; Schumpeter, 1934), social structures (Acemoglu et al., 2001; Boeke, 1953; Meier, 1964; North, 1991), human capital and R&D externalities ((Lucas, 1988; Romer, 1986), protectionism (Chang, 2008), wage differences (Emmanuel, 1972), and non-trivial ways of specialisation (Balogh, 1963; Myrdal, 1957; Singer, 1964). The FOI model therefore can reflect on the complexity of the development economics literature.

As the variegated capitalism approach warns us, interdependencies among the national economies have a crucial impact on the varieties of capitalism. A second advantage of the FOI model is that it includes indicators (trade openness, country risk, exchange rate stability) that can be interpreted as measures of these interdependencies within the world economy.

There are two practical advantages associated with the FOI analysis. Both advantages are related to the method of classification which derives groups that are comparable over time, classifies countries according to their relative performance, and limits the number of possible varieties to eight. These characteristics allow us to make a relatively complex analysis (more than two varieties) and derive stylised facts from the findings (describe typical division lines among the development paths of advanced economies). The fixed FOI structure is also useful because it makes it possible to detect country movements and model shifts over time. In other words, the hybrid model helps to check whether diversity among the OECD countries has been rising or falling, and it can be used to detect success stories.

Table 1 includes all 24 components of the FOI model. Column 3 (Factor) of Table 1 lists the variables used to calculate the 2020 FOI values. For the sake of this study, the 2010 FOI values were also required; due to some discontinued variables, a very limited number of 2010 components were calculated using a different indicator. This only causes minimal distortions because of the nature of this method. Once all factor values were collected, they were recalculated to a 1-7 scale using the min-max method (the worst factor value from all the countries included in the analysis is replaced by 1, the best one is replaced by 7, and all the others will be assigned a new value between 1 and 7). The mean of these recalculated values gives the F-, O-, and I-index of the countries. As column 1 (Potential) of Table 1 shows, the mean of the first 9 factors gives the F-index, the mean of the next 5 computes the O-index, and the mean of the rest yields the I-index.

This study is based on a comparison of OECD members. Four new members joined the OECD after 2010: Latvia, Lithuania, Colombia, and Costa Rica. These 4 countries were excluded from my analysis; therefore the total number of countries investigated is 34 (see Table 3). Once the F-, O-, and I-indices were calculated, both for 2010 and 2020, all 34 countries were placed into artificial clusters using the interval halving method. All indices are measured on a 1-7 scale, so the midpoint is 4. An index value is low (L) if it is below 4; if the value is 4 or higher, the index is considered to be high (H). With three separate indices, the number of possible combinations is 8, which means that countries can potentially be listed in 8 distinct clusters (Table 2).

Table 2. Possible clusters derived with the interval halving method (L=low, below 4; H=high)

| Cluster ID | F-index | O-index | I-index |
| --- | --- | --- | --- |
| 1 | L | L | L |
| 2 | L | L | H |
| 3 | L | H | L |
| 4 | L | H | H |
| 5 | H | L | L |
| 6 | H | L | H |
| 7 | H | H | L |
| 8 | H | H | H |

Source: own elaboration

These clusters can be interpreted as different varieties of capitalism, especially those that include a larger number of countries. The characteristics of the different versions are mainly described by the low or high values of the three indices. To obtain a deeper understanding of the clusters, I conducted a factor analysis involving more than 60 socioeconomic indicators that are correlated with at least one of the F-, O-, and I-indices. Two factors are identified in the case of each of the three pillars; the values of these factors give an insight into the inner workings of the different varieties on the one hand, and they can also explain the dynamics of the development paths within the OECD.

This study uses the following data sources:
1. OECD.Stat: https://stats.oecd.org/
2. WEF Global Competitiveness Report (Schwab 2019)
3. IMF World Economic Outlook Database, April 2021 Edition: https://www.imf.org/en/Publications/WEO/weo-database/2021/April
4. World Bank Doing Business database: https://www.doingbusiness.org/en/doingbusiness

**Results**

Table 3 presents the F-, O-, and I-indices of the 34 OECD members for 2010 and 2020. Each cell contains the index value and the rank of the country in parentheses (the index value is between 1 and 7, and the rank is between 1 and 34). Some countries excel in all three pillars (e.g., Switzerland), while others are weak in all categories (e.g., Turkey). However, most of them show mixed patterns, which is a good indication that the performance along the Future, Outside, and Inside dimensions is distinguishable.

Table 3. F-, O-, I-indices (and the rank) of the 34 OECD members for 2010 and 2020

| Country | F-2020 | F-2010 | O-2020 | O-2010 | I-2020 | I-2010 |
|---|---|---|---|---|---|---|
| Australia | 3.8 (24) | 4.6 (13) | 5.3 (4) | 5.3 (10) | 4.6 (12) | 4.4 (6) |
| Austria | 4.4 (10) | 5.1 (9) | 5.1 (8) | 5.4 (8) | 3.9 (18) | 4 (12) |
| Belgium | 3.8 (22) | 4.2 (17) | 4.9 (14) | 5.6 (5) | 3.6 (22) | 3.5 (21) |
| Canada | 4 (17) | 4.2 (18) | 4.9 (11) | 5.4 (7) | 4.6 (11) | 4.5 (2) |
| Chile | 3.6 (27) | 3.8 (21) | 3.9 (29) | 5 (14) | 3.8 (19) | 4.1 (9) |
| Czechia | 3.8 (25) | 3.4 (27) | 4.2 (25) | 5 (15) | 3.2 (25) | 3.6 (20) |
| Denmark | 4.9 (4) | 5.3 (8) | 5 (10) | 5.8 (2) | 4.7 (9) | 4.3 (7) |
| Estonia | 4.2 (16) | 3.2 (30) | 4.7 (16) | 4.9 (16) | 3.6 (21) | 3.1 (25) |
| Finland | 4.6 (7) | 5.4 (7) | 5.1 (9) | 5.7 (3) | 4.9 (6) | 4 (13) |
| France | 4.2 (15) | 4.7 (12) | 4.3 (22) | 4.5 (21) | 3.5 (23) | 3 (27) |
| Germany | 4.4 (11) | 4.8 (11) | 4.7 (17) | 5.3 (11) | 4.5 (15) | 3.7 (18) |
| Greece | 3.3 (30) | 3.1 (31) | 2.9 (34) | 3.7 (32) | 1.9 (34) | 2.5 (34) |
| Hungary | 3.1 (33) | 3.2 (29) | 4.4 (21) | 4.6 (19) | 2.6 (33) | 2.5 (33) |
| Iceland | 5.3 (1) | 5.8 (3) | 4.2 (24) | 2.3 (34) | 5 (4) | 4.4 (5) |
| Ireland | 4.3 (14) | 4.2 (19) | 4.6 (18) | 4.2 (28) | 5 (5) | 3.9 (16) |
| Israel | 4.5 (9) | 3.6 (26) | 4.6 (19) | 4.9 (17) | 4.1 (17) | 4.1 (10) |
| Italy | 3.5 (28) | 3.7 (22) | 3.5 (32) | 3.8 (30) | 2.7 (32) | 2.7 (32) |
| Japan | 4.7 (6) | 5.5 (5) | 3.7 (30) | 3.7 (31) | 4.1 (16) | 4 (14) |
| Korea | 4.3 (12) | 4.5 (14) | 4.3 (23) | 4.3 (26) | 3.8 (20) | 3.3 (22) |
| Luxembourg | 3.8 (23) | 6.1 (1) | 6.1 (1) | 6.6 (1) | 4.6 (13) | 4.5 (4) |
| Mexico | 3 (34) | 2.6 (34) | 4.1 (26) | 4 (29) | 3.3 (24) | 2.9 (30) |
| Netherlands | 4.3 (13) | 4.9 (10) | 5.3 (6) | 5.5 (6) | 5.3 (2) | 3.8 (17) |
| New Zealand | 4.5 (8) | 4.4 (15) | 5.1 (7) | 4.5 (20) | 4.8 (8) | 4 (15) |
| Norway | 4.7 (5) | 5.5 (4) | 4.9 (13) | 5.7 (4) | 4.9 (7) | 4.1 (11) |
| Poland | 3.7 (26) | 3.1 (32) | 4 (28) | 4.4 (22) | 3.1 (29) | 3.1 (26) |
| Portugal | 3.9 (19) | 3.7 (25) | 3.7 (31) | 4.3 (24) | 3.1 (28) | 2.9 (29) |
| Slovakia | 3.4 (29) | 3.3 (28) | 4.8 (15) | 4.8 (18) | 2.9 (31) | 3.3 (23) |
| Slovenia | 4 (18) | 3.7 (23) | 4.5 (20) | 5.1 (13) | 3.2 (26) | 2.7 (31) |
| Spain | 3.2 (31) | 3.7 (24) | 4 (27) | 4.2 (27) | 3.1 (27) | 3 (28) |
| Sweden | 4.9 (3) | 5.5 (6) | 4.9 (12) | 5.2 (12) | 4.6 (14) | 4.1 (8) |
| Switzerland | 5.2 (2) | 5.9 (2) | 5.4 (3) | 5.4 (9) | 5.6 (1) | 4.9 (1) |
| Turkey | 3.1 (32) | 3 (33) | 3.2 (33) | 3.6 (33) | 3.1 (30) | 3.1 (24) |
| UK | 3.8 (21) | 4.3 (16) | 5.3 (5) | 4.3 (23) | 4.7 (10) | 3.6 (19) |
| USA | 3.9 (20) | 4.1 (20) | 5.4 (2) | 4.3 (25) | 5.3 (3) | 4.5 (3) |

Source: own calculations

Applying the method of interval halving, the index values of Table 3 designate countries into different artificial clusters. The classification of the OECD countries is shown in Table 4. Based on the 2010 index values four of the possible eight clusters were described by Bartha and Gubik (2014): Cluster 1 – traditional model with outdated economic structure; Cluster 3 – dualistic model with an external resource focus; Cluster 7 – government-led, large corporation-based model with an internal safe haven focus; and Cluster 8 – human capital-based model with a focus on knowledge generation. 32 of the 34 OECD members were covered by one of these four clusters.

In 2020, the above four clusters include 29 of the 34 members. The most obvious change is the emergence of Cluster 4 (which had no members in 2010). It includes four important economies in 2020: Australia, Luxembourg, the UK, and the USA. Cluster 4 is defined by a low Future-index, and high Outside-, and Inside-indices. Except for the UK (an outlier country in 2010), all the other members were classified into this group because their F-index dropped below 4. Cluster 4 emerges as a new variety of capitalism in the FOI system; I call it the open market-based model in this study.

Although the varieties of capitalism defined in the FOI framework seem to be surprisingly stable (32 countries could be classified into 4 of them in 2010, and 33 countries into 4+1 of them in 2020), there have been significant alterations among the OECD members. Approximately half of the countries have shifted to a different cluster. Israel made the biggest positive jump, moving from a two-low – one-high (Cluster 3) group into an all-high group (Cluster 8). Estonia and Iceland have also moved into a cluster with two high index values. Most of the other countries have shifted clusters because at least one of their indices dropped below 4 (see Table 4).

Table 4. The clusters of the 34 OECD members in 2010 and 2020

| Cluster | Model name | Members in 2010 | Members in 2020 |
|---|---|---|---|
| 1 | Traditional | Greece, Italy, Mexico, Portugal, Turkey | Chile, Czechia, Greece, Italy, Poland, Portugal, Spain, Turkey |
| 2 | - | - | - |
| 3 | Dualistic | Chile, Czechia, Estonia, Hungary, Israel, Poland, Slovakia, Slovenia, Spain | Belgium, Hungary, Mexico, Slovakia, Slovenia |
| 4 | Open market-based | - | Australia, Luxembourg, UK, USA |
| 5 | - | UK | - |
| 6 | - | Iceland | Japan |
| 7 | Bureaucratic | Belgium, France. Netherlands, Ireland, Korea, New Zealand | Austria, Estonia, France, Korea |
| 8 | Human capital-based | Australia, Austria, Canada, Denmark, Finland, Germany, Japan, Luxembourg, Norway, Sweden, Switzerland, USA | Canada, Denmark, Finland, Germany, Iceland, Ireland, Israel, Netherlands, New Zealand, Norway, Sweden, Switzerland |

To better understand what lies behind the index changes, and why so many countries have shifted clusters, a factor analysis was conducted to identify the influencing factors of the FOI potentials. More than 60 socioeconomic variables were initially included in the analysis (including the components used to calculate the three indices). Six different factors were identified for the 2020 F-, O-, and I-indices (Table 5). These factors summarise the information content of several distinct variables, and their values point out the weak and strong points of a country's development path.

Table 5. The factors of the 2020 F-, O-, I-indices

| F-index factors | O-index factors | I-index factors |
|---|---|---|
| **F1 – Quality of governance** <br> Efficiency of legal framework in settling disputes <br> Property rights <br> Government ensuring policy stability <br> Strength of auditing and accounting standards <br> Population with tertiary education <br> Patent applications <br> Incidence of corruption <br> R&D expenditures <br> Life expectancy at birth <br> Contracting with Government <br> Expenditure on education | **O1 – FDI readiness** <br> Average annual wages <br> Social capital <br> Country credit rating <br> TOEFL Score <br> PISA mean score in mathematics | **I1 – Human capital** <br> Social capital <br> Ease of finding skilled employees <br> Scientific and technical journal articles <br> GDP per hour worked <br> Better life index <br> Current expenditure on health <br> Trade union coverage <br> Assets in pension funds |
| **F2 – Green growth** <br> Production-based $CO_2$ productivity <br> Emissions priced above EUR 30 per tonne of $CO_2$ <br> Renewable energy <br> Population connected to public sewerage | **O2 – Financial soundness** <br> Exchange rate stability <br> Soundness of banks | **I2 – Coordination** <br> Financing of SMEs <br> Burden of government regulation <br> Venture capital availability <br> Labor force <br> Paying Taxes <br> Labour market flexibility <br> Debt dynamics <br> Public procurement <br> Net national savings |
| Extraction Method: Principal Component Analysis <br> Rotation Method: Varimax with Kaiser Normalisation ||| 
| KMO: 0.71 <br> Bartlett's Test: Chi-Square 400.276; df 105; Sig. 0 <br> Total variance explained: 58% | KMO: 0.68 <br> Bartlett's Test: Chi-Square 107.475; df 21; Sig. 0 <br> Total variance explained: 65% | KMO: 0.80 <br> Bartlett's Test: Chi-Square 505.309; df 153; Sig. 0 <br> Total variance explained: 61% |

Principal component analysis was used as the extraction method, and Varimax with Kaiser normalisation was used as a rotation method (Table 5). The KMO values are significantly above 0.5 in all three cases, a threshold suggested to avoid larger partial correlations among the variables (Hair et al., 2018). The total variance of the F-, O-, and I-indices explained by the identified factors varies between 58% and 65%, which is not particularly high, but acceptable. The future potential of an OECD economy seems to be influenced by issues related to the quality of governance (F1), and green growth (F2). Countries with strong governance, efficient policies, and initiatives that limit emissions perform the best in this dimension of development. The two factors having an impact on a country's outside potential are issues related to the ability to absorb foreign investments (O1), and financial soundness (O2). Higher attractiveness in the eyes of foreign investors, a stable exchange rate and financial system make an OECD country score higher in the Outside index. Finally, the inside potential of the OECD countries is most dependent on the quality of human capital (I1), and the quality of coordination in allocating resources (I2). An educated and healthy labour force, and a coordination system that allocates scarce resources efficiently leads to a higher I-index value.

Table 6 shows the individual factor values for the 34 countries. Even at first sight, it is apparent that the factor values reveal even more robust individual differences than the F-, O-, and I-indices. In many cases, there are large differences between the factor pairs characterising the three different pillars: one of the factors flings the country in the positive direction (way above the OECD average), while the other one in the negative direction (below the OECD average). When evaluating the development of a single country, these factor values therefore reveal a lot more information on the path taken. Notice that some cells are empty in Table 6. This is due to the nature of the analysis: the factor cannot be calculated for a country in case of missing variables.

Table 6. Factor values for the OECD countries calculated with the 2020 F-, O-, and I-indices

| Country | F1 | F2 | O1 | O2 | I1 | I2 |
|---|---|---|---|---|---|---|
| Australia | 0.90461 | -0.68133 | 1.17784 | 1.2171 | 0.73742 | -0.05991 |
| Austria | 0.88306 | 0.66064 | 0.782 | 0.1535 | 0.38608 | 0.1122 |
| Belgium | 0.46869 | -0.42619 | 0.58483 | -0.41375 | 0.92714 | -0.77376 |
| Canada | 1.04208 | -0.99068 | 0.90789 | 0.44586 | 0.60159 | 0.30379 |
| Chile | -0.50763 | 0.7508 | -1.70085 | 2.56969 | | |
| Czechia | -0.6428 | -1.07389 | -1.19413 | 2.61477 | -1.1383 | 0.2235 |
| Denmark | 0.88487 | 0.82919 | 1.11359 | 0.03465 | 1.44666 | 0.22057 |
| Estonia | 0.02385 | -0.66533 | -0.21154 | -0.2309 | -1.80263 | 1.32682 |
| Finland | 1.61798 | 0.15286 | 0.60313 | 0.73316 | 0.71432 | 0.94483 |
| France | 0.22643 | 0.28076 | -0.07637 | -0.03617 | -0.0525 | -0.13773 |
| Germany | 0.59582 | -0.38042 | 0.80676 | -0.52349 | 0.15389 | 0.98802 |
| Greece | -1.90583 | 0.17703 | -1.87515 | -2.0794 | -0.14729 | -3.01319 |
| Hungary | -1.08498 | -0.93268 | -0.97687 | -0.89839 | -1.68816 | -0.09264 |
| Iceland | 0.30256 | 2.34261 | 0.64323 | 0.19852 | 1.8654 | -0.11877 |
| Ireland | 0.09556 | 0.89805 | 0.38738 | -0.93625 | 0.75268 | -0.09006 |
| Israel | 0.86076 | -0.41836 | -0.26673 | -0.24657 | -0.15444 | 0.46071 |
| Italy | -1.48673 | 0.69454 | -0.64899 | -1.15326 | 0.76058 | -2.84741 |
| Japan | 1.60548 | -1.44948 | -0.772 | 0.10761 | -0.99936 | 0.76482 |
| Korea | 1.14331 | -1.80278 | -0.1349 | -0.63887 | -0.61277 | -0.2097 |
| Luxembourg | 0.66251 | 1.08171 | 0.90378 | 0.41482 | 0.57606 | 0.49547 |
| Mexico | -1.41173 | -0.9826 | -1.99746 | -0.0051 | -1.16359 | -1.43735 |
| Netherlands | 0.96993 | 0.34047 | 1.19224 | 0.03595 | 0.55972 | 1.24039 |
| New Zealand | 0.67362 | 0.12666 | 0.70428 | 1.76388 | -0.36821 | 1.18571 |
| Norway | 0.46819 | 2.61942 | 0.91578 | 0.36038 | 1.45767 | 0.12275 |
| Poland | -1.06119 | -0.80835 | -0.85736 | 0.68262 | -0.83566 | -0.78477 |
| Portugal | -0.96672 | 0.51284 | -0.63504 | -1.27128 | -0.26962 | -1.14168 |
| Slovak Republic | -1.2646 | -0.36799 | -0.79413 | 0.36188 | -1.314 | -0.17608 |
| Slovenia | -0.8929 | 0.03238 | | | -0.22474 | -0.52863 |
| Spain | -0.51446 | 0.67509 | -0.46254 | -0.34132 | -0.17602 | -0.61917 |
| Sweden | 0.92475 | 0.92648 | 0.66961 | -0.65883 | 0.75981 | 0.34928 |
| Switzerland | | | | | 0.99119 | 1.11049 |
| Turkey | -1.11115 | -0.73497 | | | | |
| United Kingdom | 0.46593 | 0.76993 | 1.37484 | -1.15239 | 0.38643 | 0.55901 |
| United States | 1.12356 | -1.80388 | 1.61432 | -0.29702 | 1.10445 | 0.67575 |

Source: own calculations

**Discussion**

The hybrid approach adopted by the FOI model yielded five different varieties of capitalism within the 34 members of the OECD. Four of these were already detected in the analysis based on the 2010 data (Bartha et al., 2013), and a fifth emerged from my current investigation:

1. Traditional model: it includes countries with the worst index scores for which no clear economic policy priorities can be detected with either the F-, O-, or I-index, or with the factors of the indices. A large number of Mediterranean (or culturally similar Latin American) countries can be found in this group, which makes it similar to the Mediterranean version found by researchers such as Amable (2003) and Farkas (2016).
2. Dualistic model: countries in this cluster have low index and factor values (as in the previous group), but their outside potential is relatively high (the O-index is higher than 4), and so a common pattern in this variety is that development policy relies on external resources (mostly on foreign capital and technology). This is an export-led development strategy similar to the one suggested by Baccaro and Pontusson (2016) or Morlin et al. (2022). In a study not related to the varieties of capitalism Farooque and Khandaker (2019) also conclude that the attraction of external resources is a clear development path, and the successful implementation of this strategy has an impact on the quality of governance.
3. Government-led model: this group is the first of three that consists of countries traditionally labelled as highly developed economies. It is characterised by a high future and outside potential, and a low inside one. The factor analysis reveals that the low inside potential is largely due to lower-than-average performance in human capital (I1). To compensate for the internal vulnerability of the economy (low inside potential), these countries try to create domestic safe havens and protect the locals from international competition. The enhancing model of capitalism described by Schmidt (2005) is a good analogy for the government-led model.
4. Human capital-based model: countries in this group have high potential in all three categories. Not surprisingly, this cluster has the highest mean in human capital (I1), but the factor means are significantly above the OECD average in the other five categories as well. On the basis of the cluster members, some parallels could be drawn with the North-western and Nordic models of Farkas (2016) and the enabling model of Schmidt (2005).
5. Open market-based model: this is a cluster that emerged from the 2020 version of the FOI analysis. The four members have low future potential (besides the high outside and inside potential). The factor analysis showed that the low future potential is explained by a low mean in green growth (F2), and it also revealed that these countries are way above any other group when it comes to FDI readiness (O1). The emergence of this cluster strengthens the similarities with the results of other approaches in the field of varieties of capitalism. This cluster resembles the liberal or Anglo-Saxon model (Hall and Soskice, 2001; Amable, 2003; Schmidt, 2005). Some parallels can be drawn between the export-led, dualistic model and the open market-based one which can point to the complementarities emphasised by the variegated capitalism approach (Peck and Theodore, 2007).

The second aim of this study is to detect new patterns of development by comparing the 2010 and 2020 states of the F-, O-, and I-indices. The most obvious change is the emergence of Cluster 4 (open market-based model). Three of the countries moved from Cluster 8 to Cluster 4 because of a drop in their F-indices. The results of the factor analysis suggest that many of the moves related to future potential are due to green growth (F2) initiatives. The USA (one of the new Cluster 4 members) has the worst F2 factor value among the 34 countries; Belgium (another mover, from Cluster 7 to Cluster 3) has the worst green growth figure among all the

Western European countries. There is a clear pattern in green growth performance: 15 out of the 18 countries that have a positive factor value (a value above the OECD average) are European, and the last places (with large negative figures) are occupied by Anglo-Saxon and Asian (Korea, Japan) countries.

Climate initiatives and green growth strategies seem to have created a tear in the varieties of capitalism by 2020 that was not detectable in 2010. The barrier lies between the EU-15 (pre-2004 EU members), and all the other OECD members. There are three exceptions: Chile, New Zealand, and Norway; but Norway fits into the Nordic-Western European picture anyway. The green growth curtain spans across Europe as well: none of the countries that joined the EU in 2004 or later have a positive F2 factor value.

It is probably not a coincidence that the economies that have a tradition in coordinated markets (Hall and Soskice, 2001) and in enabling and enhancing (Schmidt, 2005) are on the positive side of the green growth curtain. The transformation initiated by climate concerns requires a lot of regulation, a lot of government intervention, and a great deal of reallocation of resources, areas in which traditional European coordinated market economies have a lot of experience.

If we concentrate on individual movers, Israel and Estonia provide two positive examples. They were both in Cluster 3 in 2010, and by 2020 Israel improved all indices to a high value (which moved it to Cluster 8), Estonia improved the F-index (and moved to Cluster 7). A closer look at the factor values of the two countries reveals that the path they took is similar. Both countries have only two positive factor values (i.e., they perform better than the OECD average in these fields), and these two factors are the same: quality of governance (F1), and coordination (I2). One interpretation of this constellation is that introducing policies and institutions that make governance and resource allocation efficient is key to modern development. Countries in cluster 3 are the ones that are traditionally regarded as being stuck in the so-called middle-income trap (Győrffy, 2022): the convergence to the most developed economies stalls following an initial growth phase that is based mostly on external resources. The FOI results suggest that these two countries managed to solve the problem of the middle-income trap by focusing on policies and institutions related to the quality of governance and coordination.

**Conclusion**

This study applies the FOI model to investigate two questions related to the varieties of capitalism: what are the typical capitalism models, or development paths among 34 OECD members; and what new patterns had emerged in the 2010-2020 period. The FOI model focuses on the future, outside and inside potential of advanced economies. These potentials are calculated by combining several components: some of them characterise norms (e.g., work ethics or social responsibility), others describe institutions (e.g., education system, government regulation), and many are related to government policies (e.g., budget expenditure in certain fields). Most of the components are supply-side factors, but demand-side elements are included, too (e.g., exports, government expenditure). The FOI model therefore represents a hybrid approach to the varieties of capitalism.

Five distinct varieties of capitalism emerge from this model. Four of these were already detected in 2010, and the 2020 analysis revealed a fifth. The traditional model (Cluster 1: Chile, Czechia, Greece, Italy, Poland, Portugal, Spain, Turkey) includes countries with low FOI potentials and does not have any other detectable features. The dualistic model (Cluster 3: Belgium, Hungary, Mexico, Slovakia, Slovenia) represents a development strategy focused on the use of external resources. The final three are similar to the enhancing, liberal, and enabling versions described by Schmidt (2005). The government-led model (Cluster 7: Austria, Estonia, France, Korea) has low inside potential, and the economies are relatively ineffective at allocating resources, which is compensated by strong government intervention. The open market-based model (Cluster 4: Australia, Luxembourg, UK, USA) emerged from the 2020 data; its future potential is low, largely due to climate regulation issues, but it is strong in the other dimensions and excels at openness and market competition. The human capital-based model (Cluster 8: Canada, Denmark, Finland, Germany, Iceland, Ireland, Israel, Netherlands, New Zealand, Norway, Sweden, Switzerland) has high FOI potentials, the highest human capital factor value, and excels at green growth.

This study also highlights two emerging patterns in the OECD members' development paths. A factor analysis showed that one of the factors having an impact on the F-index is green growth. Using the green growth factor values, this study has shown that there is a 'green growth curtain' within the OECD: the EU-15 (members that joined the EU in the 20$^{th}$ century) outperform the rest of the OECD members. It is suggested that these countries perform well in this field because (1) they have a long experience with the coordinated market model, and (2) the climate-related transition requires state regulation, intervention, and re-allocation.

Countries in clusters 1 and 3 are often interpreted as ones that are stuck in the middle-income trap. There were two of these that made a move to higher clusters: Estonia and Israel. The factor analysis showed that both countries are strong in two factors: quality of governance and coordination of resource allocation. The other message of this study is that the way out of the middle-income trap is a series of institutional and policy changes that enhance the quality of governance and make coordination more effective.